\documentclass[prc,showpacs,twocolumn,floatfix]{revtex4}
\usepackage{graphicx}
\usepackage{dcolumn}
\usepackage{bm}
\usepackage{multirow}
\usepackage{natbib}
\usepackage{color}

\begin{document}

\title{Pasta phases in neutron star studied with extended relativistic mean
field models}
\author{Neha Gupta and P.Arumugam}
\affiliation{Department of Physics, Indian Institute of Technology Roorkee,
Uttarakhand - 247 667, India}

\begin{abstract}
To explain several properties of finite nuclei, infinite matter, and neutron stars in a unified way within the relativistic mean field  models, it is important to extend them either with higher order couplings or with density-dependent couplings.  These extensions are known to have strong impact in the high-density regime.  Here we explore their role on the equation of state at densities lower than the saturation density of finite nuclei which govern the phase transitions associated with pasta structures in the crust of neutron stars.
\end{abstract}

\maketitle

\section{Introduction}
The outer layer of the neutron star (NS), with density less than the nuclear saturation density, represents different challenges and observational
opportunities such as thermal evolution, x-ray burst, glitches, and the very
important core-crust transition region \cite{Chamel10}. At this density, nucleons are correlated at short distances by attractive strong interactions- they are anticorrelated at large distances because of the Coulomb repulsion.  Competition among short- and long-range interactions leads to the development of complex and exotic nuclear shapes, which can be oversimplified \cite{ravenhall} to spheres, bubbles, rods, slabs, and tubes. These are generally
known as pasta phases. Similar geometries can also be obtained
for uncharged systems owing to finite size effects \cite{Binder:1099}.
In NS matter, geometries such as slab with holes \cite{Williams:844}, cross-rods
\cite{Pais:151101}, jungle gym,
curled spaghetti \cite{Dorso:055805}, etc., could be possible but are neglected in the present Brief Report. 
The pasta  phases in the crust  eventually dissolve into uniform matter at a certain larger density close to the saturation density. Such a critical density
at which the mantle-crust
transition
in NSs happens can be related to the neutron skin thickness in heavy nuclei
\cite{piekarewicz:0607039}.
Several other links between the NSs and finite nuclei are established \cite{FSU_para,Li:162503,ermf} and hence formulating unified models which explain both finite nuclei and NSs become important \cite{ermf}.  

The relativistic mean field (RMF) models are successful in explaining several
properties of finite nuclei. The equation of state (EoS) for symmetric
matter and NS matter obtained from many  such standard interactions do not
satisfy the experimental/observational constraints \cite{ermf,piekarewicz:0607039,DDME2}, mainly
because of their inconsistent behavior at high density.  In such
cases, it is useful to consider extensions in the RMF
models carried out by formulating an effective hadron
field theory either with additional couplings \cite{ermf,piekarewicz:0607039}-  or with density-dependent meson nucleon couplings \cite{DDME2}. These extensions do not have a strong impact on the EoS
near saturation densities (because the parameters are fitted with data for finite nuclei)  but they strongly modify the EoS at densities
higher than the saturation density.  In the present Brief Report, we analyze the
impact of the extension in the RMF models on the EoS at lower densities which
govern the occurrence of pasta phases.  

In case of higher order couplings one may relate this analysis to a trivial fact that if we fit few data points with higher order polynomials, the extrapolated results in both of the sides away from the data strongly depend on the order of the polynomial chosen. In the case of EoS, the energy density decreases with the decreasing density to reach zero at zero density, and so are the corresponding fields/interactions.  Hence, the effect of higher order couplings are not expected to be drastic but could be appreciable at finite subsaturation
densities. Pasta structures
studied with higher order couplings were first reported by us \cite{neha:dae2011d6} and a more comprehensive analysis of different nature  has recently been
published \cite{Avancini:035806}.

\section{Formalism}
In this section we briefly discuss the formalism for calculating the energy of nuclear matter in homogenous and pasta phases. For the homogenous phase, we employ RMF models  with parameters of different classes  namely,,
\begin{enumerate}
\item 
standard nonlinear interaction (NL3 \cite{nl3_para});\item density dependent interaction (DDME2 \cite{DDME2});
\item interactions with a few additional higher order couplings  (FSUGold
\cite{FSU_para} and IUFSU \cite{Fattoyev:055803}) to constrain selected observables; and
\item Effective field theory
motivated interactions  with several higher order couplings (G1 and G2 \cite{furnstahl}).
\end{enumerate}
These effective interactions are adjusted to reproduce various properties of selected finite nuclei and tested very well by explaining experimentally observed
features in a variety of nuclei. 
The details
of the Lagrangian and the coupling constants can be found in the corresponding references given above. 
The energy density for the homogenous nuclear matter is given as  
\begin{eqnarray}\epsilon&=&\sum_{i=n,p}\frac{\gamma}{(2\pi)^3}\int_{0}^{k_{fi}}d^3k\sqrt{ k^2+m_i^{*2}}-\Lambda_{v}(g_\omega
V_0)^2(g_\rho
R_0)^2\nonumber\\&&-\frac{1}{2}\left(1+\eta_1\frac{g_\sigma\sigma}{m_n}+\frac{\eta_2}{2}\frac{g_{\sigma}^2\sigma^{2}}{m_n^2}\right)m_{\omega}^{2}V^{2}_0+g_\omega
V_0(\rho_p+\rho_n)\nonumber\\&&
-\frac{1}{2}\left(1+\eta_\rho\frac{g_\sigma\sigma}{m_n}\right)m_{\rho}^{2}R_{0}^2+\frac{1}{2}g_\rho
R_0(\rho_p-\rho_n)\nonumber\\&&+{m_{\sigma}^{2}}{\sigma^{2}}\left(\frac{1}{2}+\frac{\kappa_3}{3!}\frac{g_\sigma\sigma}{m_n}+\frac{\kappa_4}{4!}\frac{g_{\sigma}^2\sigma^{2}}{m_n^2}\right)
-\frac{1}{4!}{\zeta_0}{g_{\omega}^2}V_{0}^{4}\label{ed_N},
\end{eqnarray}
where $k_{fp}=(3\pi^2x\rho)^{1/3}$, $k_{fn}=(3\pi^2(1-x)\rho)^{1/3}$, $\rho$
is the total baryon density, and $x$ defines the proton fraction. In the above expressions $\sigma$, $V_0$, and $R_0$ denote
the scalar, vector, and isovector meson mean fields, 
$m_\sigma$, $m_\omega$, and $m_\rho$ are the corresponding meson masses, and
$m_n$ is the nucleon mass. The symbols $g_\sigma$, $g_\omega$, $g_\rho$, $\kappa_3$, $\kappa_4$, $\eta_1$, $\eta_2$,
$\eta_\rho$, $\Lambda_{v}$ and $\zeta_0$ denote the various coupling constants.

\begin{figure}[t]\center
\includegraphics[width=0.99\columnwidth]{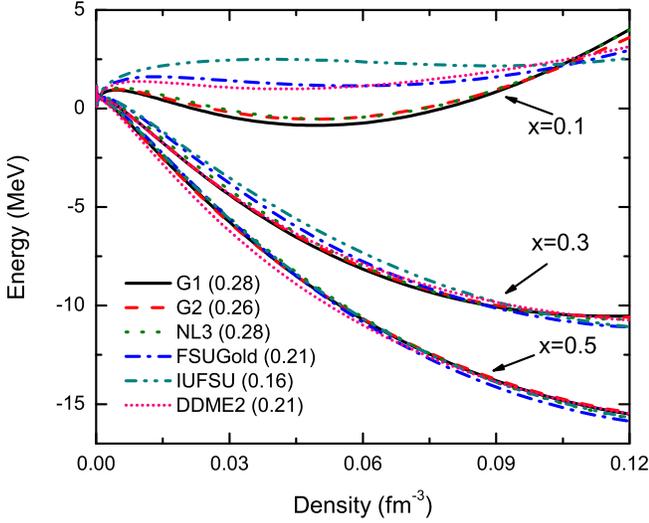}
\caption{(Color online) Energy per particle calculated with different parameter sets,
for proton fractions 0.1, 0.3, and 0.5. The numbers adjacent to the name
of the parameters represent the corresponding neutron skin thickness (in
fm) obtained
for $^{208}$Pb.}
\label{fig:EoS}
\end{figure}

\begin{figure}[t]\center
\includegraphics[width=0.8\columnwidth]{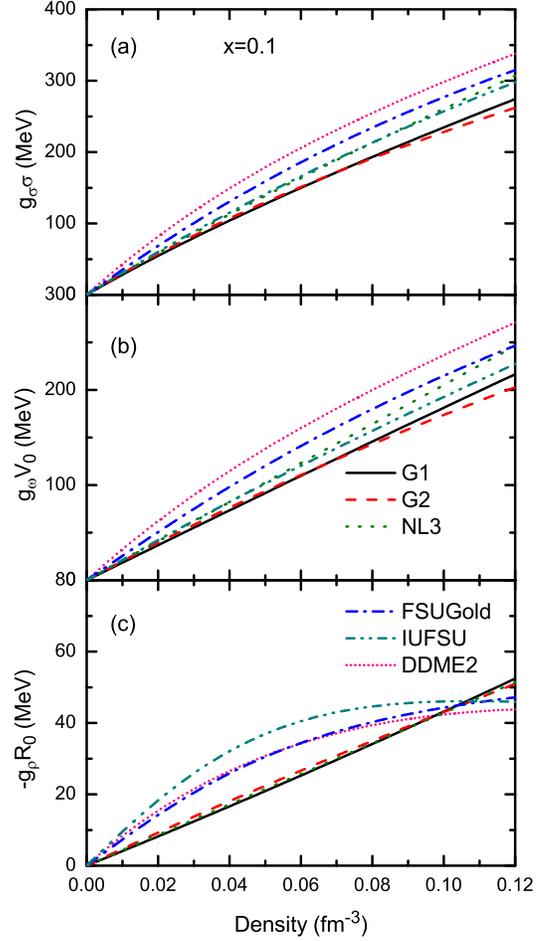}
\caption{(Color online) Density dependence of the scalar
($g_\sigma \sigma$), vector ($g_\omega V_0$), and isovector ($g_\rho R_0$) fields in the NS matter calculated with different parameter sets and at the
proton fraction $x=0.1$ .}
\label{fig:fields}
\end{figure}

\begin{figure*}[t]\center
\begin{minipage}[c]{0.75\textwidth}
\includegraphics[width=0.99\textwidth]{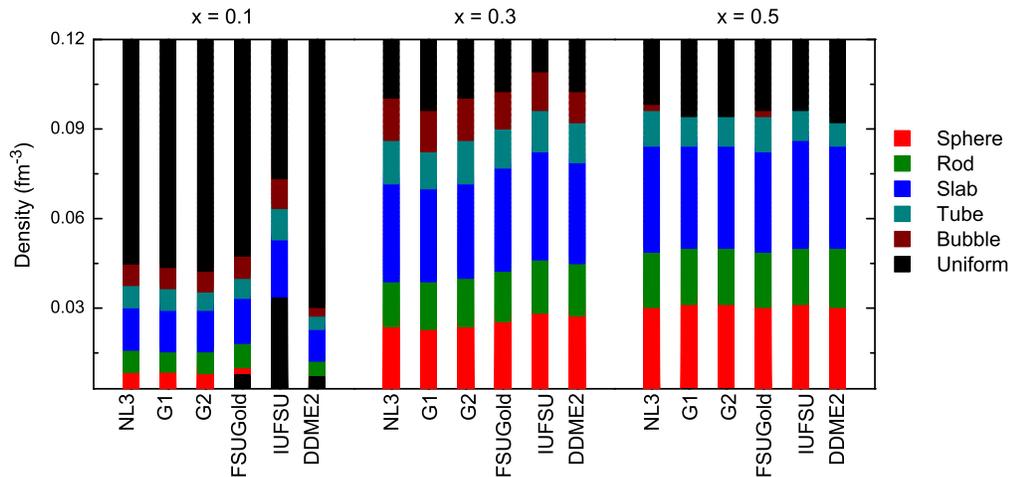}
\end{minipage} 
\begin{minipage}[c]{0.2\textwidth}
\caption{(Color online) Transition between different pasta configurations, for proton fractions 0.1, 0.3, and 0.5, calculated with different parameter sets. The different colors represent sphere, rod, slab, tube, bubble and uniform phases as given in the legend.}
\label{fig:compare}
\end{minipage} 
\end{figure*}

For the pasta phase, we employ the Wigner-Seitz approximation \cite{ravenhall} to calculate the Coulomb and surface energies for the different pasta structures.
Wigner-Seitz cells are approximated by (i) a sphere of radius $r_w$ in the case of spheres and bubbles, (ii) a cylinder of radius $r_w$ in the case of rods and tubes, and (iii) a cell bounded by planes (with $r_w$ being defined as the half distance between plane boundaries) in the case of slabs  \cite{Chamel10,ravenhall}.

Then total energy density (EoS) of the pasta phase \cite{ravenhall} is
\begin{equation}\epsilon_{tot}=\epsilon_e+u\epsilon+\epsilon_s+\epsilon_c
,\label{eq:edpasta}
\end{equation}
where $\epsilon_e$, $\epsilon_c$, and $\epsilon_s $  represent the electron, surface, and Coulomb  energy densities, respectively, and $u$ represents the  ratio of
the volume of the cluster and the Wigner-Seitz cell. In other words, $u$ represents the filling factor with $u=n/n'$, where $n$ and $n'$ represent the
average and dense phase density,
\begin{equation}\epsilon_s=u\sigma d/r  \text{ and } \epsilon_c=2u(e^{2}\pi\phi)^{1/3}(\sigma
dx n'/4)^{2/3},
\end{equation}
where $u=u$ for droplets and $u=1-u$ for bubbles, $\sigma$ is the surface
tension coefficient calculated by a Thomas-Fermi method \cite{Grill:055808}, $d=1,2,3$ represents  the dimension of the system, and 
\begin{equation}
\phi=\frac{1}{(d+2)}\left[\frac{2}{(d-2)}\left(1-\frac{1}{2}du^{1-2/d}\right)+u\right].
\end{equation} 
In the case of $d=2$, the above expression can be rewritten as 
\begin{equation}
\phi=\frac{(u-1)-\log u}{4}.
\end{equation} 

The radius of the droplet (rod, slab) and that of the Wigner-Seitz cell are,
respectively, given by
\begin{equation}
r=\phi^{-1/3}(4\pi n'^2x^2 e^2/(\sigma d))^{-1/3}, r_w=r/u^{1/d}.
\end{equation}

\section{Results and discussion}

We first examine the variation of energy at subsaturation densities relevant
for pasta phase. Figure~\ref{fig:EoS} represents the energy per particle for nuclear matter as a function of density calculated with different
parameter sets and proton fractions. We can see the general trend that with increasing proton fraction the  matter is becoming more and more bound. For symmetric nuclear matter (proton fraction $x=0.5$)  the binding is maximum and we do not see much difference in the results from different parameters. The differences start to appear at $x=0.3$ and with $x=0.1$, we can identify three clear groups of parameter sets: (i) having a shallow well in the energy (G1,G2, NL3), (ii) saturated energy at low density (FSUGold, DDME2), and (iii) with hump in the energy at low density (IUFSU). To understand the above grouping and related features in terms of the different fields and their couplings, we look into the density dependence of the $\sigma$, $\omega$, and $\rho$ fields, which are shown in Fig.~\ref{fig:fields}.  We notice that at subsaturation
densities, the $\sigma$ field is the most dominant. The higher order terms
in the $\sigma$ field (self-coupling) are taken care of in all the chosen parameters
upto the fourth order.  DDME2 is an exception with linear $\sigma$ field
but with explicitly density-dependent couplings which decrease drastically
with density for the range shown in Fig.~\ref{fig:fields}. The $\omega$ field follows the same pattern of the $\sigma$ field and hence we may not be able to see the explicit role of higher order couplings involving the $\omega$ field alone or those involving both $\sigma$ and $\omega$.  

We see quite interesting features in the $\rho$ field which is almost linear in Fig.~\ref{fig:fields}(c) for NL3, G1 and G2. The difference in energy owing to the change in $\rho$ field could be expected through the second, fifth and sixth terms in Eq.~(\ref{ed_N}). Among these terms the second and fifth terms can contribute to the nonlinear behavior of $\rho$ field (while $\sigma$
and $\omega$ fields are almost linear and the proton fraction is constant).
Between these two terms, the second one (with the coupling constant $\Lambda_v$) dominates but this term appears exclusively in the parameters FSUGold and IUFSU.  Thus the higher order coupling $\Lambda_v$, representing the second order $\omega$-$\rho$ interaction, plays a major role and we can identify that
the hump in the energy is proportional to $\Lambda_v$. This parameter plays an important role at higher densities also, as evident from its influence on NS mass and radius \cite{neha}. 

\begin{figure*}[t]\center
\begin{minipage}[c]{0.68\textwidth}
\includegraphics[width=0.95\textwidth]{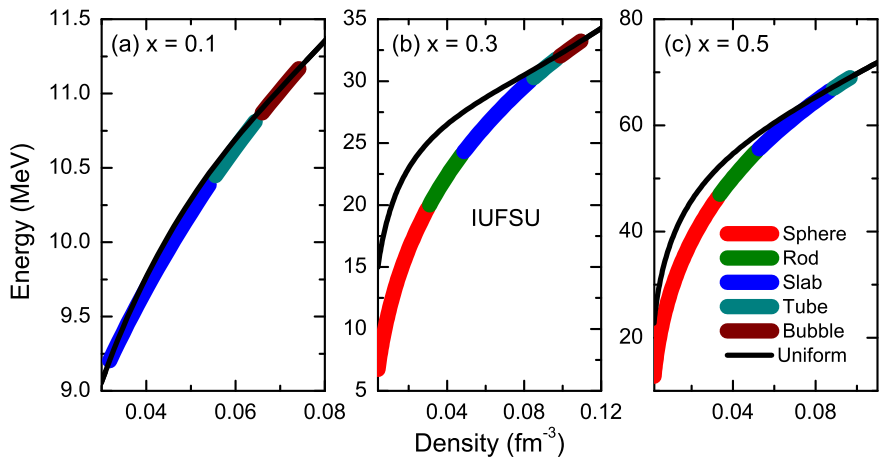}
\end{minipage} 
\begin{minipage}[c]{0.28\textwidth}
\caption{(Color online) Energy per particle for the uniform matter and
pasta-like matter obtained with the IUFSU parameter set for proton fraction
(a) 0.1, (b) 0.3, and (c) 0.5. The different colors represent the different pasta structures
as mentioned in the inset.  The energy contribution from the electrons is
also included here (unlike Fig.~1).}
\label{fig:IUFSU2}
\end{minipage} 
\end{figure*}

The parameter set DDME2, even without the higher order coupling, yields a
$\rho$ field similar to that of FSUGold because the coupling
constant $g_\rho$ varies nonlinearly with density \cite{DDME2} and mimics
the role of $\Lambda_v$.  It has to be noted that the neutron
skin thickness has been considered to be an ingredient in the fitting procedure
of the three
parameters DDME2, FSUGold and IUFSU. The values of neutron skin thickness
obtained for $^{208}$Pb are also shown in Fig.~\ref{fig:EoS} along with the legends.
We can see that the ordering of energy at $x=0.1$ for different parameters
is same as that of the neutron skin thicknesses.  As the neutron skin thickness
is strongly correlated to the density-dependence of symmetry energy, we can
associate the differences in energies to the differences in the symmetry
energies.  Hence it is more appropriate to link the low-density behavior
of EoS (especially for neutron rich case) to the symmetry energy as discussed
in Ref.~\cite{Grill:055808}. The surface tension coefficients \cite{Grill:055808} also follow
the same ordering of the energy and neutron skin thickness. In Fig.~1, we can see that the different parameters converge
at a density lower than the saturation density ($\approx 0.11$ fm$^{-3}$
for $x=0.1$).  This is attributable to the fact that the parameters are fitted to
reproduce the properties of finite nuclei whose average density is lower
than the saturation density \cite{Khan:092501}.

We identify that the $\rho$ field is the major source of the difference
in the EoS from different parameters.  At higher proton fractions the $\sigma$
and $\omega$ fields remain almost the same and the strength of the $\rho$ field decreases. Thus, the results from different parameters
converge at higher proton fraction.
      
The various phase transitions happening in the NS comprising the considered
pasta structures, obtained from our calculations with different proton fractions and parameter
sets, are represented in Fig.~\ref{fig:compare}. It has to be noted that
there could be more geometries of pasta structures \cite{Pais:151101,Dorso:055805,Williams:844} and hence Fig.~\ref{fig:compare} is
not a complete phase diagram. With all the parameter
sets, we find almost the same trend in the sequence of pasta structures. At low density, sphere phase is the most stable configuration, followed by
rod, slab, tube, and bubble phases. In most of the cases, a homogeneous matter phase occurs at a density $\sim0.11$ fm$^{-3}$, which, incidentally, is the
point where all the EoS converge \cite{Khan:092501}. The difference between results
from different parameter sets could be clearly linked to the features discussed
above in the case of EoS [Fig.~1]. Similar to the case of energy, at higher proton fractions the different parameter sets yield similar
results and we see interesting features at $x= 0.1$. The sustained existence
of uniform phase at lower densities in the cases of FSUGold, IUFSU and DDME2
can be linked to the corresponding  EoS shown in Fig.~\ref{fig:EoS}, which are larger and raising. The energy difference between uniform phase and pasta phase is very small at $x= 0.1$, and such a delicate balance allows the differences in EoS to show up significantly by altering the transition densities from one phase to other.

In Fig~4, we compare the energies calculated for the uniform phase and the
pasta phases, in case
of IUFSU for which the difference with other parameters is notable in Fig.~3.
The crossing between these energies determine the corresponding phase transitions.
The
main difference between the uniform phase energy in Fig.~\ref{fig:EoS} and Fig.~\ref{fig:IUFSU2} is that in Fig.~\ref{fig:IUFSU2} we include the
contribution from electron energy also. In general, the pasta phase energy [Eq.(\ref{eq:edpasta})] is dominated by the corresponding uniform phase energy.  At $x=0.1$, the contribution from the Coulomb term is lesser and hence the difference between energies of pasta phases and uniform phase is smaller, as evident from Fig.~4.  At larger proton fractions the Coulomb
term widens the energy gap and the difference between different EoSs starts
to diminish, leading to similar phase transitions.

\section{Conclusions}
With lower proton fractions, the higher order couplings can play a role in modifying the EoS at subsaturation densities relevant for the pasta structures in neutrons
stars. The $\sigma$ field dominates the EoS at subsaturation density and
its higher order couplings are taken care of all the realistic models.  The fourth-order coupling between $\omega$ and $\rho$ fields ($\Lambda_v$), plays a crucial role at lower proton fraction and a similar role is played by the
density-dependent $\rho$ field coupling constant. At higher proton fractions,
the $\rho$ field, which is the source of the difference between different
parameter sets, gets weaker and hence all the results converge. We observe
that it is more appropriate to link
the differences in EoS (and hence the occurrence of pasta structures) to the symmetry energy (and hence the neutron skin thickness).


\begin{thebibliography}{19}
\expandafter\ifx\csname natexlab\endcsname\relax\def\natexlab#1{#1}\fi
\expandafter\ifx\csname bibnamefont\endcsname\relax
  \def\bibnamefont#1{#1}\fi
\expandafter\ifx\csname bibfnamefont\endcsname\relax
  \def\bibfnamefont#1{#1}\fi
\expandafter\ifx\csname citenamefont\endcsname\relax
  \def\citenamefont#1{#1}\fi
\expandafter\ifx\csname url\endcsname\relax
  \def\url#1{\texttt{#1}}\fi
\expandafter\ifx\csname urlprefix\endcsname\relax\def\urlprefix{URL }\fi
\providecommand{\bibinfo}[2]{#2}
\providecommand{\eprint}[2][]{\url{#2}}

\bibitem[{\citenamefont{Chamel and Haensel}(2008)}]{Chamel10}
\bibinfo{author}{\bibfnamefont{N.}~\bibnamefont{Chamel}} \bibnamefont{and}
  \bibinfo{author}{\bibfnamefont{P.}~\bibnamefont{Haensel}},
  \bibinfo{journal}{Living Reviews in Relativity} \textbf{\bibinfo{volume}{11}}
  (\bibinfo{year}{2008}).

\bibitem[{\citenamefont{Ravenhall et~al.}(1983)\citenamefont{Ravenhall,
  Pethick, and Wilson}}]{ravenhall}
\bibinfo{author}{\bibfnamefont{D.~G.} \bibnamefont{Ravenhall}},
  \bibinfo{author}{\bibfnamefont{C.~J.} \bibnamefont{Pethick}},
  \bibnamefont{and} \bibinfo{author}{\bibfnamefont{J.~R.}
  \bibnamefont{Wilson}}, \bibinfo{journal}{Phys. Rev. Lett.}
  \textbf{\bibinfo{volume}{50}}, \bibinfo{pages}{2066} (\bibinfo{year}{1983}).

\bibitem[{\citenamefont{Binder et~al.}(2012)\citenamefont{Binder, Block,
  Virnau, and Tr{\"o}ster}}]{Binder:1099}
\bibinfo{author}{\bibfnamefont{K.}~\bibnamefont{Binder}},
  \bibinfo{author}{\bibfnamefont{B.~J.} \bibnamefont{Block}},
  \bibinfo{author}{\bibfnamefont{P.}~\bibnamefont{Virnau}}, \bibnamefont{and}
  \bibinfo{author}{\bibfnamefont{A.}~\bibnamefont{Tr{\"o}ster}},
  \bibinfo{journal}{Am. J. Phys.} \textbf{\bibinfo{volume}{80}},
  \bibinfo{pages}{1099} (\bibinfo{year}{2012}).

\bibitem[{\citenamefont{Williams and Koonin}(1985)}]{Williams:844}
\bibinfo{author}{\bibfnamefont{R.}~\bibnamefont{Williams}} \bibnamefont{and}
  \bibinfo{author}{\bibfnamefont{S.}~\bibnamefont{Koonin}},
  \bibinfo{journal}{Nuclear Physics A} \textbf{\bibinfo{volume}{435}},
  \bibinfo{pages}{844 } (\bibinfo{year}{1985}).

\bibitem[{\citenamefont{Pais and Stone}(2012)}]{Pais:151101}
\bibinfo{author}{\bibfnamefont{H.}~\bibnamefont{Pais}} \bibnamefont{and}
  \bibinfo{author}{\bibfnamefont{J.~R.} \bibnamefont{Stone}},
  \bibinfo{journal}{Phys. Rev. Lett.} \textbf{\bibinfo{volume}{109}},
  \bibinfo{pages}{151101} (\bibinfo{year}{2012}).

\bibitem[{\citenamefont{Dorso et~al.}(2012)\citenamefont{Dorso,
  Gim\'enez~Molinelli, and L\'opez}}]{Dorso:055805}
\bibinfo{author}{\bibfnamefont{C.~O.} \bibnamefont{Dorso}},
  \bibinfo{author}{\bibfnamefont{P.~A.} \bibnamefont{Gim\'enez~Molinelli}},
  \bibnamefont{and} \bibinfo{author}{\bibfnamefont{J.~A.}
  \bibnamefont{L\'opez}}, \bibinfo{journal}{Phys. Rev. C}
  \textbf{\bibinfo{volume}{86}}, \bibinfo{pages}{055805}
  (\bibinfo{year}{2012}).

\bibitem[{\citenamefont{Piekarewicz}(2006)}]{piekarewicz:0607039}
\bibinfo{author}{\bibfnamefont{J.}~\bibnamefont{Piekarewicz}},
 in Proceedings of the International Conference on Current Problems of Nuclear Physics and Atomic Energy (NPAE-2006), 29 May–3 June 2006, Kyiv, Ukraine, edited by I. M. Vyshnevskyi and V. P. Verbytskyi (the Institute for Nuclear Research of National Academy of Sciences, Ukraine, 2007), pp. 33–42.

\bibitem[{\citenamefont{Todd-Rutel and Piekarewicz}(2005)}]{FSU_para}
\bibinfo{author}{\bibfnamefont{B.~G.} \bibnamefont{Todd-Rutel}}
  \bibnamefont{and}
  \bibinfo{author}{\bibfnamefont{J.}~\bibnamefont{Piekarewicz}},
  \bibinfo{journal}{Phys. Rev. Lett.} \textbf{\bibinfo{volume}{95}},
  \bibinfo{pages}{122501} (\bibinfo{year}{2005}).

\bibitem[{\citenamefont{Li et~al.}(2007)\citenamefont{Li, Garg, Liu, Marks,
  Nayak, Rao, Fujiwara, Hashimoto, Kawase, Nakanishi et~al.}}]{Li:162503}
\bibinfo{author}{\bibfnamefont{T.}~\bibnamefont{Li}},
  \bibinfo{author}{\bibfnamefont{U.}~\bibnamefont{Garg}},
  \bibinfo{author}{\bibfnamefont{Y.}~\bibnamefont{Liu}},
  \bibinfo{author}{\bibfnamefont{R.}~\bibnamefont{Marks}},
  \bibinfo{author}{\bibfnamefont{B.~K.} \bibnamefont{Nayak}},
  \bibinfo{author}{\bibfnamefont{P.~V.~M.} \bibnamefont{Rao}},
  \bibinfo{author}{\bibfnamefont{M.}~\bibnamefont{Fujiwara}},
  \bibinfo{author}{\bibfnamefont{H.}~\bibnamefont{Hashimoto}},
  \bibinfo{author}{\bibfnamefont{K.}~\bibnamefont{Kawase}},
  \bibinfo{author}{\bibfnamefont{K.}~\bibnamefont{Nakanishi}},
  \bibnamefont{et~al.}, \bibinfo{journal}{Phys. Rev. Lett.}
  \textbf{\bibinfo{volume}{99}}, \bibinfo{pages}{162503}
  (\bibinfo{year}{2007}).

\bibitem[{\citenamefont{Arumugam et~al.}(2004)\citenamefont{Arumugam, Sharma,
  Sahu, Patra, Sil, Centelles, and Vi\~nas}}]{ermf}
\bibinfo{author}{\bibfnamefont{P.}~\bibnamefont{Arumugam}},
  \bibinfo{author}{\bibfnamefont{B.~K.} \bibnamefont{Sharma}},
  \bibinfo{author}{\bibfnamefont{P.~K.} \bibnamefont{Sahu}},
  \bibinfo{author}{\bibfnamefont{S.~K.} \bibnamefont{Patra}},
  \bibinfo{author}{\bibfnamefont{T.}~\bibnamefont{Sil}},
  \bibinfo{author}{\bibfnamefont{M.}~\bibnamefont{Centelles}},
  \bibnamefont{and} \bibinfo{author}{\bibfnamefont{X.}~\bibnamefont{Vi\~nas}},
  \bibinfo{journal}{Phys. Lett. B} \textbf{\bibinfo{volume}{601}},
  \bibinfo{pages}{51 } (\bibinfo{year}{2004}).

\bibitem[{\citenamefont{Lalazissis et~al.}(2005)\citenamefont{Lalazissis,
  Nik\ifmmode \check{s}\else \v{s}\fi{}i\ifmmode~\acute{c}\else \'{c}\fi{},
  Vretenar, and Ring}}]{DDME2}
\bibinfo{author}{\bibfnamefont{G.~A.} \bibnamefont{Lalazissis}},
  \bibinfo{author}{\bibfnamefont{T.}~\bibnamefont{Nik\ifmmode \check{s}\else
  \v{s}\fi{}i\ifmmode~\acute{c}\else \'{c}\fi{}}},
  \bibinfo{author}{\bibfnamefont{D.}~\bibnamefont{Vretenar}}, \bibnamefont{and}
  \bibinfo{author}{\bibfnamefont{P.}~\bibnamefont{Ring}},
  \bibinfo{journal}{Phys. Rev. C} \textbf{\bibinfo{volume}{71}},
  \bibinfo{pages}{024312} (\bibinfo{year}{2005}).

\bibitem[{\citenamefont{Gupta et~al.}(2011)\citenamefont{Gupta, Shabnam, and
  Arumugam}}]{neha:dae2011d6}
\bibinfo{author}{\bibfnamefont{N.}~\bibnamefont{Gupta}},
  \bibinfo{author}{\bibfnamefont{I.}~\bibnamefont{Shabnam}}, \bibnamefont{and}
  \bibinfo{author}{\bibfnamefont{P.}~\bibnamefont{Arumugam}}, in
  \emph{\bibinfo{booktitle}{Proceedings of the DAE Symposium on Nuclear
  Physics}} (\bibinfo{year}{2011}), vol.~\bibinfo{volume}{56}, p.
  \bibinfo{pages}{702},
  \urlprefix\url{http://www.sympnp.org/proceedings/56/D6.pdf}.

\bibitem[{\citenamefont{Avancini et~al.}(2012)\citenamefont{Avancini, Barros,
  Brito, Chiacchiera, Menezes, and Provid\^encia}}]{Avancini:035806}
\bibinfo{author}{\bibfnamefont{S.~S.} \bibnamefont{Avancini}},
  \bibinfo{author}{\bibfnamefont{C.~C.} \bibnamefont{Barros}},
  \bibinfo{author}{\bibfnamefont{L.}~\bibnamefont{Brito}},
  \bibinfo{author}{\bibfnamefont{S.}~\bibnamefont{Chiacchiera}},
  \bibinfo{author}{\bibfnamefont{D.~P.} \bibnamefont{Menezes}},
  \bibnamefont{and}
  \bibinfo{author}{\bibfnamefont{C.}~\bibnamefont{Provid\^encia}},
  \bibinfo{journal}{Phys. Rev. C} \textbf{\bibinfo{volume}{85}},
  \bibinfo{pages}{035806} (\bibinfo{year}{2012}).

\bibitem[{\citenamefont{Lalazissis et~al.}(1997)\citenamefont{Lalazissis,
  K\"onig, and Ring}}]{nl3_para}
\bibinfo{author}{\bibfnamefont{G.~A.} \bibnamefont{Lalazissis}},
  \bibinfo{author}{\bibfnamefont{J.}~\bibnamefont{K\"onig}}, \bibnamefont{and}
  \bibinfo{author}{\bibfnamefont{P.}~\bibnamefont{Ring}},
  \bibinfo{journal}{Phys. Rev. C} \textbf{\bibinfo{volume}{55}},
  \bibinfo{pages}{540} (\bibinfo{year}{1997}).

\bibitem[{\citenamefont{Fattoyev et~al.}(2010)\citenamefont{Fattoyev, Horowitz,
  Piekarewicz, and Shen}}]{Fattoyev:055803}
\bibinfo{author}{\bibfnamefont{F.~J.} \bibnamefont{Fattoyev}},
  \bibinfo{author}{\bibfnamefont{C.~J.} \bibnamefont{Horowitz}},
  \bibinfo{author}{\bibfnamefont{J.}~\bibnamefont{Piekarewicz}},
  \bibnamefont{and} \bibinfo{author}{\bibfnamefont{G.}~\bibnamefont{Shen}},
  \bibinfo{journal}{Phys. Rev. C} \textbf{\bibinfo{volume}{82}},
  \bibinfo{pages}{055803} (\bibinfo{year}{2010}).

\bibitem[{\citenamefont{Furnstahl et~al.}(1997)\citenamefont{Furnstahl, Serot,
  and Tang}}]{furnstahl}
\bibinfo{author}{\bibfnamefont{R.~J.} \bibnamefont{Furnstahl}},
  \bibinfo{author}{\bibfnamefont{B.~D.} \bibnamefont{Serot}}, \bibnamefont{and}
  \bibinfo{author}{\bibfnamefont{H.-B.} \bibnamefont{Tang}},
  \bibinfo{journal}{Nucl. Phys. A} \textbf{\bibinfo{volume}{615}},
  \bibinfo{pages}{441 } (\bibinfo{year}{1997}).

\bibitem[{\citenamefont{Grill et~al.}(2012)\citenamefont{Grill, Provid\^encia,
  and Avancini}}]{Grill:055808}
\bibinfo{author}{\bibfnamefont{F.}~\bibnamefont{Grill}},
  \bibinfo{author}{\bibfnamefont{C.}~\bibnamefont{Provid\^encia}},
  \bibnamefont{and} \bibinfo{author}{\bibfnamefont{S.~S.}
  \bibnamefont{Avancini}}, \bibinfo{journal}{Phys. Rev. C}
  \textbf{\bibinfo{volume}{85}}, \bibinfo{pages}{055808}
  (\bibinfo{year}{2012}).

\bibitem[{\citenamefont{Gupta and Arumugam}(2012)}]{neha}
\bibinfo{author}{\bibfnamefont{N.}~\bibnamefont{Gupta}} \bibnamefont{and}
  \bibinfo{author}{\bibfnamefont{P.}~\bibnamefont{Arumugam}},
  \bibinfo{journal}{Phys. Rev. C} \textbf{\bibinfo{volume}{85}},
  \bibinfo{pages}{015804} (\bibinfo{year}{2012}).

\bibitem[{\citenamefont{Khan et~al.}(2012)\citenamefont{Khan, Margueron, and
  Vida\~na}}]{Khan:092501}
\bibinfo{author}{\bibfnamefont{E.}~\bibnamefont{Khan}},
  \bibinfo{author}{\bibfnamefont{J.}~\bibnamefont{Margueron}},
  \bibnamefont{and} \bibinfo{author}{\bibfnamefont{I.}~\bibnamefont{Vida\~na}},
  \bibinfo{journal}{Phys. Rev. Lett.} \textbf{\bibinfo{volume}{109}},
  \bibinfo{pages}{092501} (\bibinfo{year}{2012}).

\end{thebibliography}
\end{document}